# Large Band Gap Quantum Anomalous Hall Phase in Hexagonal Organometallic Frameworks

Yuanjun Jin[1], Zhongjia Chen[1,2], Bowen Xia[1], Yujun Zhao[2], Rui Wang[3,*], Hu Xu[1,†]

[1] *Department of Physics & Institute for Quantum Science and Engineering, Southern University of Science and Technology, Shenzhen 518055, P. R. China.*
[2] *Department of Physics, South China University of Technology, Guangzhou 510640, P. R. China*
[3] *Institute for Structure and Function & Department of physics, Chongqing University, Chongqing 400044, P. R. China.*

The nontrivial band gap plays a critical role in quantum anomalous Hall (QAH) insulators. In this work, we propose that the intrinsic QAH phase with sizable band gaps up to 367 meV is achieved in two-dimensional hexagonal organometallic frameworks (HOMFs). Based on first-principles calculations and effective model analysis, we uncover that these large band gaps in transition metal based HOMFs are opened by strong spin-orbital coupling of the localized *4d* or *5d* electrons. Importantly, we reveal that Coulomb correlations can further significantly enhance the nontrivial band gaps. In addition, we suggest a possible avenue to grow these organometallic QAH insulators on a semiconducting SiC(111) substrate, and the topological features are perfectly preserved due to the van der Waals interaction. Our work shows that the correlation remarkably enhances the nontrivial band gaps, providing exotic candidates to realize the QAH states at high temperatures.

The quantum anomalous Hall (QAH) state was first proposed in a two-dimensional (2D) lattice model with spontaneously time-reversal symmetry breaking by Haldane in 1988 [1]. The nontrivial band topology of a QAH system is characterized by a topological invariant $C$ known as the first Chern number. According to the global topology of Bloch states, the Chern number $C$ is defined as the integral of the Berry curvature over the occupied bands in the entire Brillouin zone (BZ) [2]. As a result, the chiral edge states associated with the quantized Hall conductance in QAH insulators are topologically protected and robustly stable against both magnetic and nonmagnetic impurities. This intriguing property provides potentially significant applications in future nanoelectronics and spintronics. Hence, extensive efforts have always been devoted in exploring realistic QAH candidates. Up to now, there have been numerous 2D materials theoretically predicted to possess QAH effects, such as graphene/magnetic-insulator heterostructures [3-6], heavy element layers [7], semi-functionalized stanene and bismuthene[8], and transition metal (TM) doped topological insulators (TIs) [9-12]. In addition, several TM based hexagonal organometallic frameworks (HOMFs) [13-15] have also been proposed to possess QAH effects since these organometallic compounds can be possibly synthesized using chemical solution or substrate-mediated self-assembled growth methods[16-18]. Unfortunately, it is challenging to experimentally realize the QAH states in realistic materials due to the stringent requirements. So far, QAH effects were only observed in Cr or V doped (Bi, Sb)$_2$Te$_3$ thin films at extremely low temperatures (~30 meV) [9,10,12,19]. As it is quite difficult to prepare doped (Bi, Sb)$_2$Te$_3$ films with a long-range magnetic ordering, the comprehensive study of QAH effects is seriously hindered. Therefore, it is crucially important to explore promising QAH candidates that host large band gaps and high Curie temperatures ($T_C$).

In this work, based on first-principles calculations and effective model analysis, we propose that 2D HOMFs bridged by TM atoms (TM=Ta, Nb) can satisfy these above criteria. In comparison with previously reported organometallic QAH insulators with *3d* electrons [13,14], our proposed HOMFs with Ta *5d* or Nb *4d* electrons possess greater intrinsic SOC strength. More importantly, the *d* orbitals of TM atoms in these HOMFs exhibit strong Anderson localization. Therefore, Coulomb correlations significantly enhance nontrivial band gaps [20-24], resulting in sizable band gaps up to 367 meV, which are considerably larger than those reported in other HOMFs [13,14]. Different from the interaction-driven Dirac system, in which the quantum fluctuations strongly suppress the QAH states and its ground state would be more likely to be driven into other competing orders [25,26], nontrivial band gaps in our proposed HOMEs are driven by SOC rather than interaction. This is essentially the same with QAH states in other Dirac systems [13,14,27]. Furthermore, a semiconducting SiC(111) substrate can be chosen to epitaxially grow Ta based HOMFs. The corresponding nontrivial features are perfectly preserved due to the van der Waals interaction, suggesting the possible experimental realization of these promising QAH states.

We performed density functional theory (DFT) calculations as implemented in Vienna *ab initio* Simulation Package [28]. The projector augmented wave method were used [29]. The exchange correlation functional was treated within the generalized gradient approximation in Perdew-Burke-Ernzerbof (PBE) formalism [30]. Due to the strong localization of *d* orbitals in HOMFs, the Heyd-Scuseria-Ernzerhof (HSE06) hybrid functional was adopted to correct the Coulomb correlations [31]. The vacuum space was set to 18 Å to avoid the artificial interaction between neighboring layers. The BZ was sampled by using a 6×6×1 Monkhorst-Pack grid. The cutoff energy of plane wave basis was set to 500 eV and the forces on each atom were smaller than 0.01 eV/Å. The phonon dispersion was calculated within the force constant approach as implemented in the PHONOPY code [32]. The tight-binding (TB) Hamiltonian was constructed by using maximally localized Wannier functions (MLWF) methods by using the WANNIER90 package [33].

As shown in Fig. 1(a), a primitive unit cell of HOMFs consists of two Ta atoms and three benzene rings, i.e., the chemical formula is $Ta_2C_{18}H_{12}$. Three neighboring benzene rings are connected by one Ta atom, forming a 2D hexagonal lattice with unit cell vectors $a_1$ and $a_2$, and $|a_1|=|a_2|=a$. Due to the strong steric repulsion, benzene rings prefer to rotate around the Ta-Ta axis, and Ta atoms move alternately up and down out of the plane. As a result, a buckled geometry with a vertical height $h$ is present. The optimized lattice parameters for $TM_2C_{18}H_{12}$ (TM=Ta, Nb) are listed in Table 1. In the main text, we pay attention to the results of $Ta_2C_{18}H_{12}$, and detailed results of $Nb_2C_{18}H_{12}$ are included in the Supplemental Material (SM) [34]. The phonon dispersion is one useful way to characterize the stability and structural rigidity. Take $Ta_2C_{18}H_{12}$ as an example, the corresponding phonon dispersion is shown in Fig. 1(b). We can see that there are no soft modes, suggesting the kinetical stability of $Ta_2C_{18}H_{12}$. Besides, we also performed *ab initio* molecular dynamics (AIMD) simulations to further confirm their thermodynamic stabilities, and the corresponding results are shown in Figs. S1 and S2 [34].

Next, we consider the magnetic properties of $TM_2C_{18}H_{12}$ (TM=Ta, Nb). To study the energetically favorable magnetic ground states, we carried out calculations with different magnetic configurations. The results show that the total energies of ferromagnetic (FM) states are lower than those of antiferromagnetic (AFM) states, and the FM ground states possesses a spontaneous magnetic moment of $2\mu_B$ localized on each TM atom. Furthermore, magnetic anisotropy calculations indicate that the direction of spin polarization is vertical to the $TM_2C_{18}H_{12}$ plane (i.e., along the *z*-direction). The out-of-plane FM configuration of $Ta_2C_{18}H_{12}$ is 21 meV per unit cell lower than that of the in-plane FM configuration. Because TM (TM=Ta, Nb) atoms belong to the group VB and share the similar *d* orbital shells, the origin of the magnetic moment of $2\mu_B$ can be understood in the following scheme. Each TM atom bonds to three carbon atoms, and then there are two unpaired $d$ electrons left. According to the Hund's rule, each TM atom has the same spin $S = 1$, thus rendering $2\mu_B$ per each TM atom. Furthermore, we performed AIMD simulations at 400 K to investigate the magnetic thermodynamics of $TM_2C_{18}H_{12}$ (TM = Ta, Nb). As shown in Figs. S1 and S2 [31], we can see that their FM properties are perfectly reserved during 10 ps AIMD simulations, suggesting that the magnetic configurations of our proposed HOFMs are robust above the room temperature.

We can use the Ising model to estimate the Curie temperatures of $TM_2C_{18}H_{12}$. The spin Hamiltonian can be described by $H = -\sum_{\langle i,j \rangle} JS_iS_j$, where $J$ represents the nearest-neighbor exchange parameter, and $S_i$ is the magnetic moment of TM atoms [35]. For a hexagonal lattice [36], the exchange parameter $J = (E_{AFM} - E_{FM})/6S^2$ is adopted, where $E_{FM} = -3JS^2$ and $E_{AFM} = 3JS^2$. The calculated exchange parameters $J$ are 12.67 meV and 10.70 meV for Ta and Nb, respectively. Then, we carried out Monte Carlo (MC) simulations to calculate the Curie temperature $T_C$, and a $80 \times 80$ supercell with the periodic boundary condition is employed. The results of MC simulations are shown in Fig. 1(c). The calculated Curie temperatures are 900 K for $Ta_2C_{18}H_{12}$ and 720 K for $Nb_2C_{18}H_{12}$, which are much higher than the room temperature.

In the following, we focus on the electronic and topological features of $Ta_2C_{18}H_{12}$, and the results of $Nb_2C_{18}H_{12}$ are included in Figs. S3 and S4 [34]. As shown in Fig. 2(a), the projected densities of states (PDOS) in the absence of SOC indicate that the states near the Fermi level are dominated by the majority spin states of Ta $d_{xz}$ and $d_{yz}$ orbitals, while the contributions from other orbitals are negligible. Without SOC, the spin-polarized band structures of $Ta_2C_{18}H_{12}$ respectively obtained from PBE and

HSE06 calculations are shown in Fig. 2(b). We find that the majority spin (blue line) and minority spin (red line) bands are completely separated from each other near the Fermi level. The minority spin channel exhibits an insulating feature. Remarkably, two majority spin bands exactly cross at the Fermi level at the $K$ (or $K'$) points in the hexagonal BZ. This Dirac half-metallic semimetal is protected by $C_3$ rotational symmetry. In the presence of SOC, the Dirac cone must be destroyed due to the spin-rotation symmetry breaking [37], and then the Dirac half-metallic state converts into the QAH insulating state. As expected, PBE calculations give a band gap of ~138 meV for $Ta_2C_{18}H_{12}$ by taking into account of SOC [the left panel of Fig. 2(c)]. The band gap of $Ta_2C_{18}H_{12}$ is considerably larger than those of previously reported organometallic QAH insulators with *3d* orbitals. This is because Ta *5d* orbitals possess the greater SOC strength. Furthermore, it is worth noting that the large distance between Ta atoms in $Ta_2C_{18}H_{12}$ lead to the strong localization of *d* orbitals. To more accurately treat correlation effects of the localized electrons, HSE06 calculations with SOC are carried out, and the corresponding band structure is shown in the right panel of Fig. 2(c). Remarkably, one finds that the nontrivial band gap at the $K$ point is significantly enhanced to 367 meV.

As mentioned above, the band gap induced by the SOC is nontrivial. To confirm the topological properties, we calculated Berry curvature $\Omega(\mathbf{k})$ of $Ta_2C_{18}H_{12}$ over the all occupied states in 2D momentum space as shown in Fig. 2(d). The nonzero $\Omega(\mathbf{k})$ diverges at the $K$ (or $K'$) points with the same sign and vanishes away from the $K$ (or $K'$) points. By the integral of the Berry curvatures over the occupied bands in the whole BZ, we obtain that the first Chern number $C$ is equal to -1, revealing the presence of QAH states. One hallmark of a QAH insulator is the chiral edge state. Based on MLWF methods, we constructed a TB Hamiltonian with Green's function method to obtain the edge states using WANNIERTOOLS package [38]. In Fig. 3(a), we plot the energy spectrum of a semi-infinite $Ta_2C_{18}H_{12}$ ribbon terminated with zigzag edge. The chiral edge state connecting the valence and conduction bands is clearly visible in the bulk band gap. We also calculated the intrinsic Hall conductance from the Kubo formula using the Wannier TB Hamiltonian [33,39]. As expected, the Hall conductance inside the band gap exhibits the quantized plateau ($\sigma_{xy} = e^2/h$) [see Fig. 3(b)].

To insightfully capture the topological mechanism of the correlation-enhanced large band gap QAH states in $Ta_2C_{18}H_{12}$, we construct an effective model based on the projected-orbital analysis. As shown in Fig. 2(a), we can work with a two-orbital model including Ta $d_{xz}$ and $d_{yz}$ orbitals with $i, j = 1(|A, xz\rangle)$, $2(|A, yz\rangle)$, $3(|B, xz\rangle)$, $4(|B, yz\rangle)$, where A and B denote the sublattice indices. In this Wannier representation, the majority spin Hamiltonian can be written as [22]

$$H = \sum_{i,j,\mathbf{R}} \left( t_{ij} d_i^\dagger d_j + H.c. \right) + i\lambda \sum_{i,j,\mathbf{R}} (\mathbf{E}_{ij} \times \mathbf{R}_{ij}) \cdot \mathbf{s} d_i^\dagger d_j + U \sum_{\mathbf{R}} \left( \hat{n}_{A,xz} \hat{n}_{A,yz} + \hat{n}_{B,xz} \hat{n}_{B,yz} \right), \quad (1)$$

where $d_i^\dagger$ ($d_i$) creates (annihilates) an electron with basis $i$, $\mathbf{E}_{ij}$ is the electric field from neighboring sites experienced along $\mathbf{R}_{ij}$, $\mathbf{s}$ is the spin Pauli matrix, and $U$ is the effective on-site inter-orbital correlation parameter. In Eq. (1), the first term is the nearest-neighbor TB Hamiltonian ($H_0$) constructed from PBE calculations without SOC and the hopping parameters $t_{ij}$ can be obtained from the MLWF method [33,39]. We compare the band structure obtained from the TB Hamiltonian with the result from first-principles calculations in Fig. S5 [34], showing that the band structure of considering the nearest-neighbor hopping agrees well with the DFT result around the $K$ (or $K'$) point. The second term describes SOC ($H_{SOC}$) with the magnitude $\lambda$ [27]. The third term is on-site Coulomb inter-orbital correlation ($H_{int}$) with the density operator $\hat{n}_i = d_i^\dagger d_i$ [21]. Using the Hartree-Fock approximation $\langle \hat{n}_i \rangle = n_i$, the on-site correlation $H_{int}$ is decoupled and we can obtain its band structure by diagonalizing the Hamiltonian in Eq. (1) in reciprocal space. In order to clearly reveal the correlation-enhanced band gap in $Ta_2C_{18}H_{12}$, we show the band gap as a function of $U$ in Fig. 4(a). We can see that the band gap of $Ta_2C_{18}H_{12}$ rapidly increases with increasing the correlation blocking $U$, exhibiting a strong correlation-enhanced effect on the nontrivial band gap. When $U = 2.4$ eV, the band gap of $Ta_2C_{18}H_{12}$ reaches to 367 meV, which is the value calculated from the hybrid functional calculation with $\alpha = 0.25$ of the exact exchange. We also change the mixture parameter $\alpha$ in hybrid functional calculations, and the corresponding results are shown in Fig. 4(b). Our analysis definitely shows that the Coulomb correlation and the fraction of the exact exchange play a similar role in the correlation effect of $5d$ orbitals. In addition, we also investigate the SOC effects on the band gap. Figure 4(c) shows the variation of band gap with $U = 2.4$ eV as a function of the SOC strength $\lambda$ in the range of 0 ~ 1. Starting from the calculation without SOC ($\lambda = 0$), it can be seen that the band gap nearly linearly increases with increasing $\lambda$. Therefore, we conclude that these large band gaps in our proposed $TM_2C_{18}H_{12}$ (TM= Ta, Nb) are opened by strong SOC of the localized *4d* or *5d* electrons, and Coulomb correlations further enhance nontrivial band gaps.

It is well known that it is challenging to synthesis freestanding QAH insulators due to their poor thermodynamic stability. 2D organometallics may be grown on metal substrates, since metal may act as a proper catalyst for metal atoms and molecules to assemble [18]. However, the metal substrate will bring in trivial bands near the Fermi level and make the nontrivial features being difficultly detectable. The investigations also indicated that the growth of 2D topological insulators on semiconductors involving strong chemical bonding is feasible in experiments [27,40,41]. If the realistic QAH films are epitaxially grow on a semiconducting substrate, their nontrivial features remain intact. Considering the symmetry and lattice mismatch, we suggest that the hydrogen-terminated SiC(111) surface is possible substrate for epitaxial growth of our proposed HOMFs. As shown in Fig. 5(a), $Ta_2C_{18}H_{12}$ is deposited on the SiC(111) surface to construct a $Ta_2C_{18}H_{12}$/SiC(111) heterostructure. The lattice mismatch between $Ta_2C_{18}H_{12}$ and the

$p(3\times3)$ SiC(111) surface is only 2.3%. The optimized distance between $Ta_2C_{18}H_{12}$ and the SiC(111) surface is found to be 2.64 Å with the interfacial binding energy 0.085 $J/m^2$, suggesting that the $Ta_2C_{18}H_{12}$/SiC(111) interface is a typical van der Waals (vdW) heterostructure and improves the thermodynamic stability of $Ta_2C_{18}H_{12}$. We have carried out AIMD simulations to further confirm the thermodynamic stability of this vdW heterostructure. The corresponding results are shown Fig. S6 [34]. We can see that this structure is stable at least up to 400 K. The projected band structure without SOC is shown in Fig. 5(b). As expected, the Dirac cone at the $K$ (or $K'$) point is preserved, which is consistent with the freestanding case. These results indicate that the SiC(111) surface can be served as a possible substrate to epitaxially grow our proposed HOMFs with desirable topological features.

In summary, based on first-principles calculations and effective model analysis, we proposed that $TM_2C_{18}H_{12}$ (TM = Ta, Nb) are intrinsic QAH insulators with sizable nontrivial band gaps up to ~367 meV, which are originated from the interplay between the Coulomb correlation and strong SOC. The nontrivial features are confirmed by the chiral edge states associated with the quantized Hall conductance. Furthermore, we find that a semiconducting SiC(111) substrate offers a possible avenue for epitaxial growth of the proposed $Ta_2C_{18}H_{12}$ while their topological features can be perfectly preserved. Our findings not only exhibit a promising topological mechanism with correlation-enhanced nontrivial band gaps, but also provide exotic candidates to realize the QAH states at high temperatures.

*E-mail: rcwang@cqu.edu.cn (R.W.)

†E-mail: xuh@sustc.edu.cn (H.X)

**ACKNOWLEDGMENTS**

This work was supported by the National Natural Science Foundation of China (NSFC, Grant Nos. 11674148, 11334004, 91634106, and 11404159), the Guangdong Natural Science Funds for Distinguished Young Scholars (No. 2017B030306008), the Basic Research Program of Science, Technology, Innovation Commission of Shenzhen Municipality (Grant Nos. JCYJ20160531190054083, JCYJ20170412154426330), the Fundamental Research Funds for the Central Universities of China (Grant Nos. 106112017CDJXY300005 and cqu2018CDHB1B01), and Science Challenge Project (No. TZ2016003).

Table I. The lattice constant $a$, buckled height $h$ of two nonequivalent TM (TM = Ta, Nb) atoms, PBE band gap ($\Delta_{PBE}$), HSE06 band gap ($\Delta_{HSE}$), and energy difference between AFM and FM states of $TM_2C_{18}H_{12}$.

| TM | $a$ (Å) | $h$ (Å) | $\Delta_{PBE}$ (meV) | $\Delta_{HSE}$ (meV) | $E_{AFM} - E_{FM}$ (meV) |
|---|---|---|---|---|---|
| Ta | 12.085 | 0.58 | 138 | 367 | 304 |
| Nb | 12.092 | 0.84 | 32 | 278 | 257 |

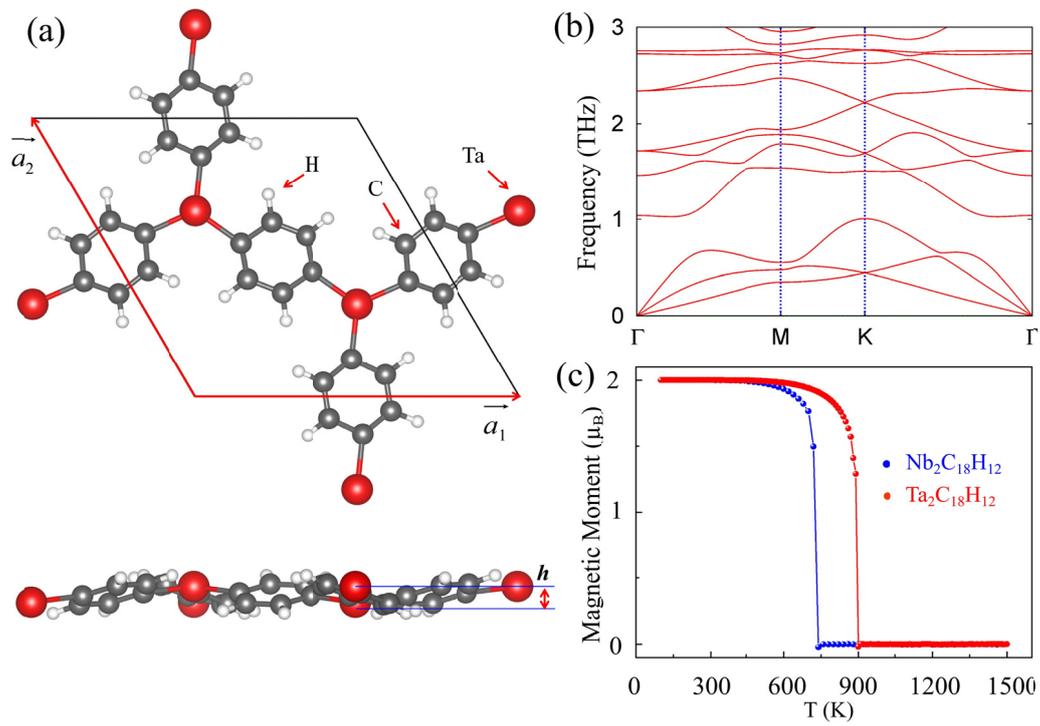

Figure 1 (a) Top and side views of monolayer $Ta_2C_{18}H_{12}$. The unit cell is marked by a hexagonal lattice with unit cell vectors $a_1$ and $a_2$. The bucked height of two nonequivalent Ta atoms is labeled by $h$. (b) The calculated phonon dispersion of $Ta_2C_{18}H_{12}$ with the ferromagnetic configuration. (c) MC simulations of $T_C$ of $TM_2C_{18}H_{12}$ (TM = Ta, Nb).

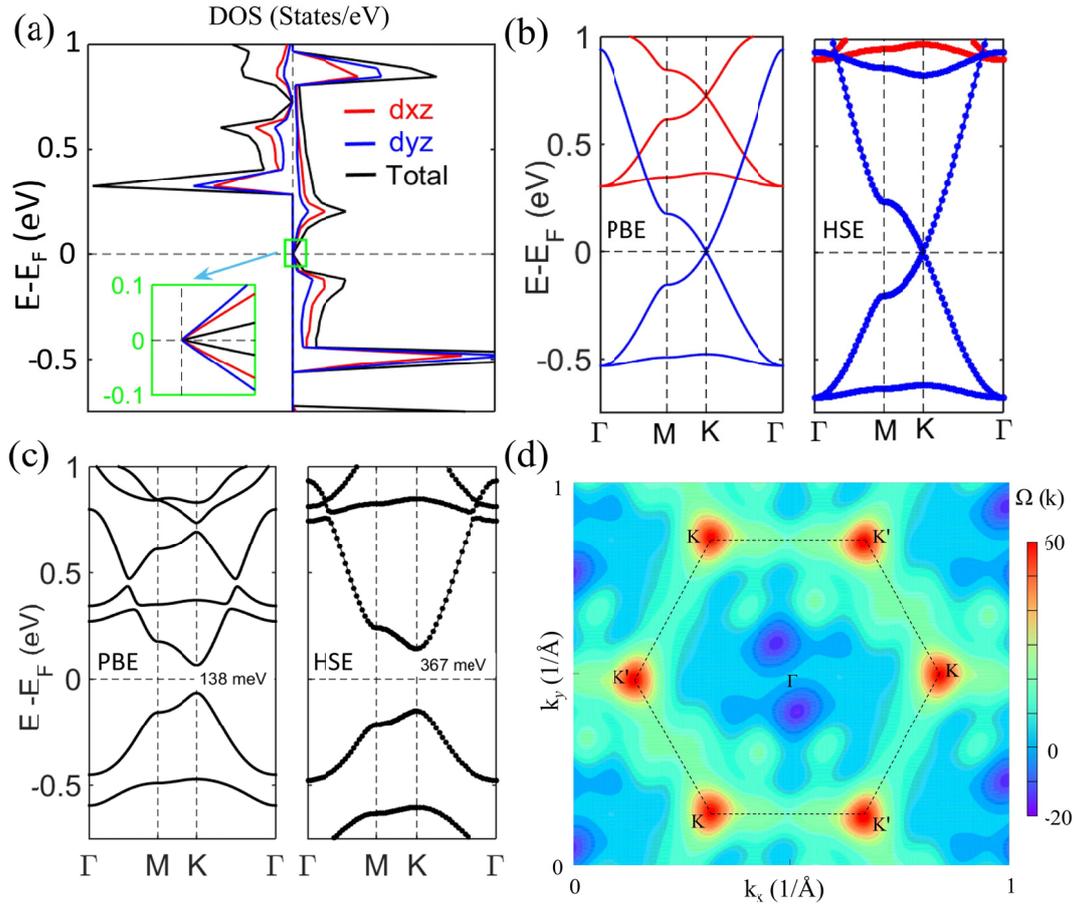

Figure 2 (a) The spin-polarized PDOS obtained by PBE calculations. (b) The spin-polarized band structures of Ta$_2$C$_{18}$H$_{12}$ obtained by PBE (left) and HSE06 (right) calculations in the absence of SOC. The blue and red lines denote the majority and minority spin bands, respectively. (c) The PBE (left) and HSE06 (right) band structures of Ta$_2$C$_{18}$H$_{12}$ in the presence of SOC. (d) The distribution of the Berry curvature $\Omega(\mathbf{k})$ in the presence of SOC. The first BZ is marked by the black dashed lines.

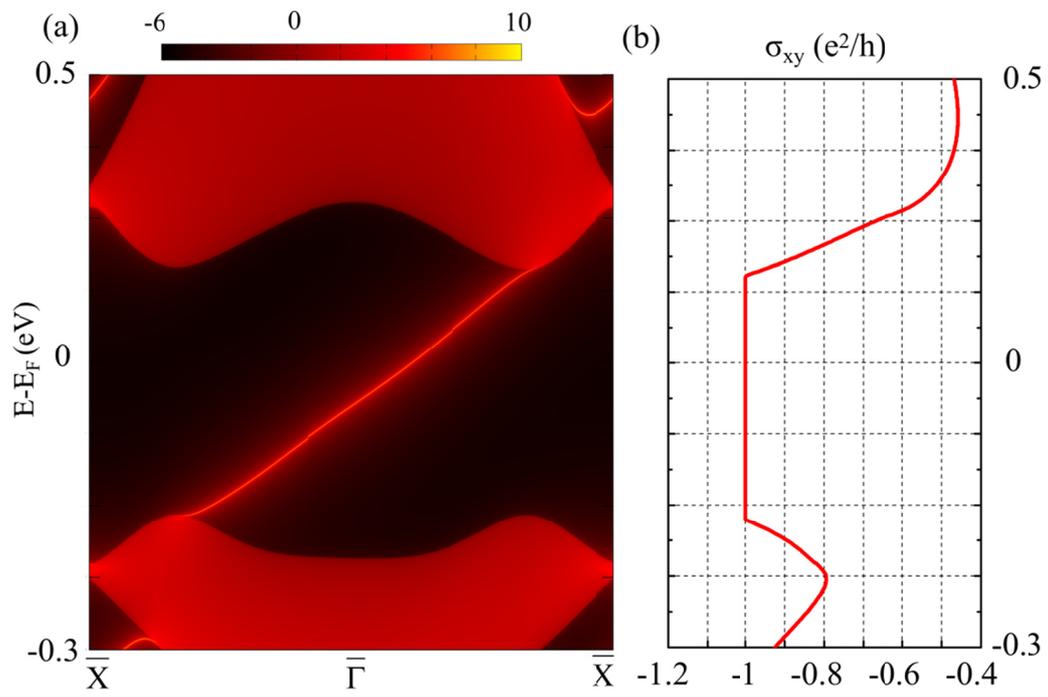

Figure 3 (a) TB band structure of a semi-infinite $Ta_2C_{18}H_{12}$ ribbon terminated by its zigzag edge. The chiral edge state is clearly visible inside the bulk gap. (b) The calculated anomalous Hall conductance $\sigma_{xy}$ in units of $e^2/h$.

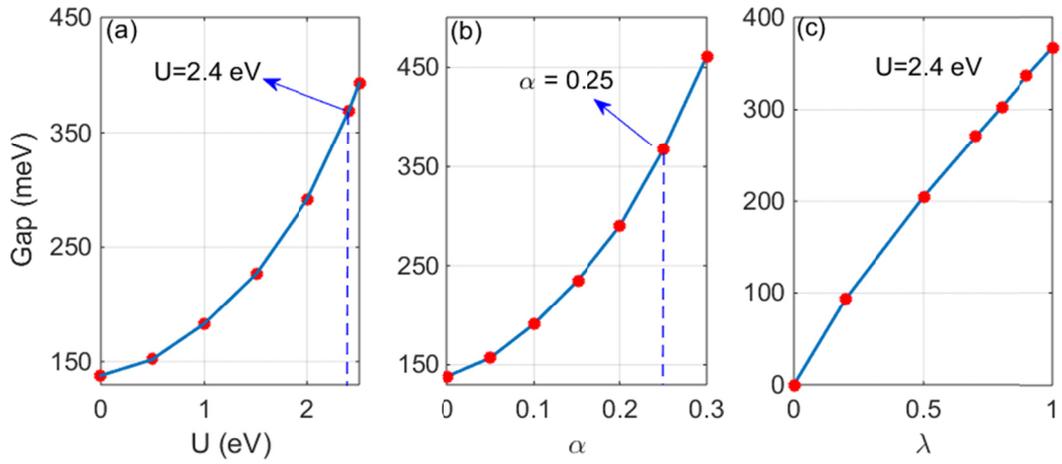

Figure 4 (a) Band gaps as a function of $U$ of $Ta_2C_{18}H_{12}$, and the band gap of 367 meV at $U = 2.4$ eV is labeled. (b) Band gaps of $Ta_2C_{18}H_{12}$ as a function of mixture fraction of the exact exchange $\alpha$ in hybrid functional calculations. (c) The variation of band gap of $Ta_2C_{18}H_{12}$ with $U = 2.4$ eV as a function of the SOC strength $\lambda$ in the range of 0 ~ 1.

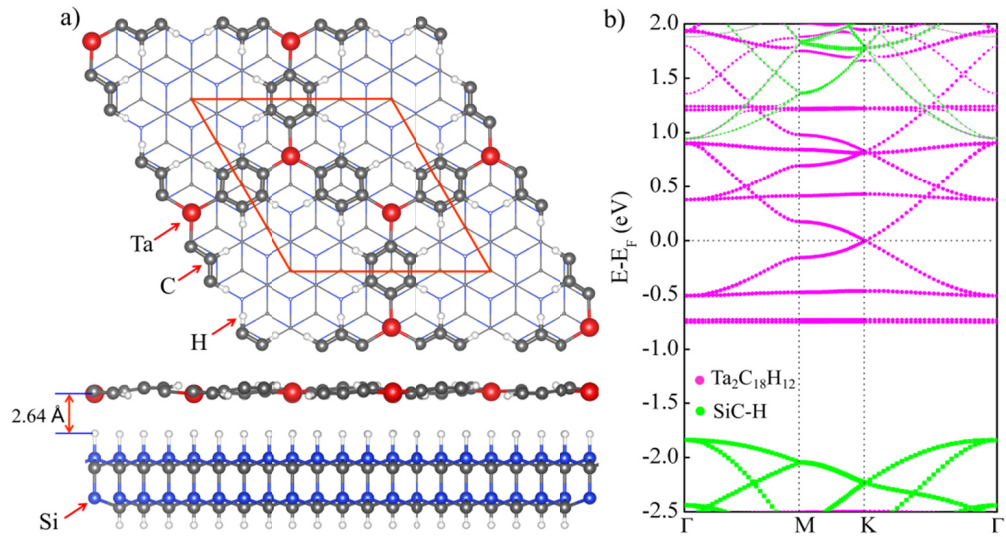

Figure 5 (a) Top and side views of the $Ta_2C_{18}H_{12}$/SiC(111) heterostructure. The unit cell of $Ta_2C_{18}H_{12}$ is highlighted by red solid lines. (b) The projected band structure for the $Ta_2C_{18}H_{12}$/SiC(111) heterostructure in the absence of SOC. The Dirac cone at the K (or K') point is preserved. The band structures projected on $Ta_2C_{18}H_{12}$ and the SiC(111) substrate are plotted by pink and green lines, respectively.